\documentclass[conference]{IEEEtran}
\IEEEoverridecommandlockouts

\usepackage{cite}
\usepackage{amsmath,amssymb,amsfonts}
\usepackage{algorithm}
\usepackage{algpseudocode}
\usepackage{graphicx}
\usepackage{textcomp}
\usepackage{xcolor}
\usepackage{multirow}
\usepackage{epstopdf}
\usepackage{multicol}
\usepackage[symbol]{footmisc}
\usepackage{float}
\usepackage{tikz}

\def\BibTeX{{\rm B\kern-.05em{\sc i\kern-.025em b}\kern-.08em
    T\kern-.1667em\lower.7ex\hbox{E}\kern-.125emX}}
\begin{document}

\title{UX-aware Rate Allocation for Real-Time Media}

\author{\IEEEauthorblockN{Belal Korany, Peerapol Tinnakornsrisuphap, Saadallah Kassir, Prashanth Hande, \\Hyun Yong Lee, and Thomas Stockhammer}
\IEEEauthorblockA{\textit{Qualcomm Technologies, Inc.} \\
San Diego, CA, USA \\
\{bkorany,peerapol,skassir,phande,hyunyong,tsto\}@qti.qualcomm.com}
}

\maketitle

\begin{abstract}
Immersive communications is a key use case for 6G where applications require reliable latency-bound media traffic at a certain data rate to deliver an acceptable User Experience (UX) or Quality-of-Experience (QoE). The Quality-of-Service (QoS) framework of current cellular systems (4G and 5G) and prevalent network congestion control algorithms for latency-bound traffic like L4S typically target network-related Key Performance Indicators (KPIs) such as data rates  and latencies. Network capacity is based on the number of users that attain these KPIs. However, the UX of an immersive application for a given data rate and latency is not the same across users, since it depends on other factors such as the complexity of the media being transmitted and the encoder format. This implies that guarantees on network KPIs do not necessarily translate to guarantees on the UX. 

In this paper, we propose a framework in which the communication network can provide guarantees on the UX. The framework requires application servers to share real-time information on UX dependency on data rate to the network, which in turn, uses this information to maximize a UX-based network utility function. Our framework is motivated by the recent industry trends of increasing application awareness at the network, and pushing application servers towards the edge, allowing for tighter coordination between the servers and the 6G system. Our simulation results show that the proposed framework substantially improves the UX capacity of the network, which is the number of users above a certain UX threshold, compared to conventional rate control algorithms.
\end{abstract}


\section{Introduction}
Recently, there has been a rapid growth in the applications and deployment of eXtended Reality (XR) technologies, which encapsulate Virtual Reality (VR), Augmented Reality (AR), and Mixed Reality (MR). The use cases and demands for these technologies are still expected to grow exponentially, with VR being forecasted to be a US\$62B market by 2027  \cite{amiri2024application}. These technologies create immersive user experiences (which is adopted by ITU-R as a key use case for 6G \cite{recommendation2023framework}) and pose challenging requirements on cellular networks, such as high data rates and very tight latency budgets. In order to accommodate for these requirements, 3GPP has introduced several enhancements in the 5G system (5GS), e.g., better Quality-of-Service (QoS) handling, 5GS information exposure, and application awareness at the network \cite{hande2023extended}. On the QoS front, 5GS introduced the support of PDU-set-based QoS handling, where a PDU-set is a term used to represent a collection of packets that carry a single media unit, e.g. a video frame. On the network information exposure front, 5GS adopted Explicit Congestion Notification (ECN) marking for the support of Low Latency, Low Loss, and Scalable Throughput (L4S) traffic \cite{3gpp.23.501}. This means that a 5G network node (e.g., RAN) can mark some IP packets to quickly notify applications of congestion conditions, which helps with rate adaptation at the application layer. As for application awareness, an application may be able to provide the network with some PDU-set information (e.g., periodicity, jitter, size, ...) through PDU-set metadata or through standalone assistance information, which helps the network manage its resources \cite{amiri2024application}. 

\begin{figure}
\begin{center}
\includegraphics[width=0.9\linewidth]{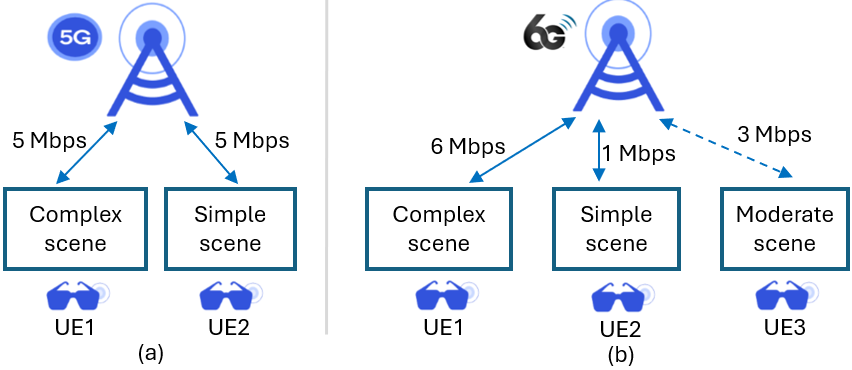}
\caption{Scenario of interest: (a) Two UEs with similar channel conditions will get similar network resources, despite their different video complexities. (b) A UX-aware rate allocation can improve the network's performance. }
\label{fig:motivation}
\end{center}
\vspace{-13pt}
\end{figure}

All the aforementioned enhancements rely on network-level Key Performance Indicators (KPIs), including data rates and latencies, to measure the network's performance and optimize its operation. Further, the network capacity is measured in terms of the number of simultaneous users who meet these network-level KPIs. These KPIs and frameworks, however, are inadequate for immersive applications since they fail to characterize the User Experience (UX). For example, any network-centric rate adaptation framework (such as L4S), rely on a coarse assumption that higher throughput for a UE equals better UX, without any notion of how much any UE can benefit from getting extra network resources \cite{nadas2024qoe}. 

To better clarify the issue, consider the example in Fig.~\ref{fig:motivation}~(a), where two XR devices are connected to the same 5G cell. At one point in time, the first device (UE1) is streaming a very complex video with a lot of spatial and temporal details, while the second (UE2) is streaming a simple video. If the channel qualities of the two devices are similar, the two UEs will end up sharing the network resources equally and getting similar data rates, negatively impacting the experience of UE1 and not materially improving that of UE2. This is due to the fact that the 5G network has no awareness of the media content, and that the two flows, both being XR flows, are assigned the same QoS level. On the other hand, if the network were aware of the media content complexities, then the same video quality could have been delivered for UE2 with a much lower bitrate, freeing up resources to boost the bitrate of UE1, or even to accommodate the addition of other UEs to the network and increase the UX-capacity, defined as the number of simultaneous users that meet a UX threshold, see Fig.~\ref{fig:motivation}~(b). 

Another issue with conventional congestion control algorithms is their dependence on End-to-End (E2E) application feedback for rate adaptation. Closing this loop typically requires tens of milliseconds, making the response to sudden channel variations slow, when compared to the tight latency requirements of XR traffic.

In this paper, we build on the existing trend of increasing application awareness at the access network and propose a framework in which the Application Server (AS) shares \textit{real-time media complexity information} with the network, and the network shares \textit{direct rate allocation feedback} to the AS. This kind of fast information sharing becomes more feasible with the recent trend of pushing ASs towards the edge, which will result in easier and faster coordination between the AS and the network's components.  We show the benefits of UX-awareness at the network by proposing two possible rate allocation algorithms which maximize different UX-based network-utility functions: (1) maximizing UX or Quality-of-Experience (QoE)\footnote{As QoE is a metric for measuring UX, we use the terms interchangeably for the rest of the paper.} capacity, and (2) maximizing minimum QoE. Our simulation results show that the proposed framework leads to significant gains in terms of both application-level KPIs such as UE satisfaction, and network-level KPIs such as E2E latency. This presents a paradigm shift for how cellular networks handle the requirements of different data flows: \textit{from QoS to QoE}.

The importance and possible benefits of UX awareness at the network level has been recognized by few recent papers in the research community \cite{nadas2024qoe,yan2022qoe,slivar2019qoe,liebl2005radio}. In \cite{nadas2024qoe}, the authors recognize the issue that different video streams have different complexities, and that the complexity of a single video stream may vary drastically over time. They propose a resource sharing algorithm that takes this issue into account, albeit, without rigorous validation for the algorithm's performance. In \cite{yan2022qoe}, the authors propose a QoE-aware resource allocation algorithm for semantic communications, where the QoE model is developed for task-oriented information delivery over the network, which is not suitable for XR traffic. The authors of \cite{slivar2019qoe} also propose a QoE aware rate allocation framework. However, in their model, each cloud game (or category of games) has one constant time-invariant QoE value, which is not the case for realistic XR traffic.

\section{User Experience (UX) model}
\label{sec:qoe_model}
The UX model depends on the media type. This section will discuss a model based on real-time video streaming for XR services which is characterized by very tight latency requirements. To meet this requirement, minimal buffering is implemented on the Application Server (AS) or Application Client (AC), and the media frames are immediately transmitted from the AS with minimal delay. This results in a periodic traffic pattern (with some jitter) whose bursts (frames) have an average size of the current bitrate of the application encoder divided by the frame rate. To avoid queue build-up at the network, the application's bitrate needs to be continuously adapted to varying channel conditions and network congestion.

To characterize the UX of a video stream, several QoE metrics have been proposed in the literature \cite{min2024perceptual}. These metrics can be broadly classified into two categories: 
\subsubsection{Temporal quality} describing the smoothness of video playback. When a frame is not delivered in time for the device display, the decoder copies the last successfully decoded frame to the display, and the video is said to be in a \textit{stall}. Temporal quality can be measured by the AC, using metrics such as the Maximum Stall Duration (MSD) and stall frequency, which are both functions of the tail of the frame latencies.
\subsubsection{Spatial quality} 
\begin{figure}
\begin{center}
\includegraphics[width=1\linewidth]{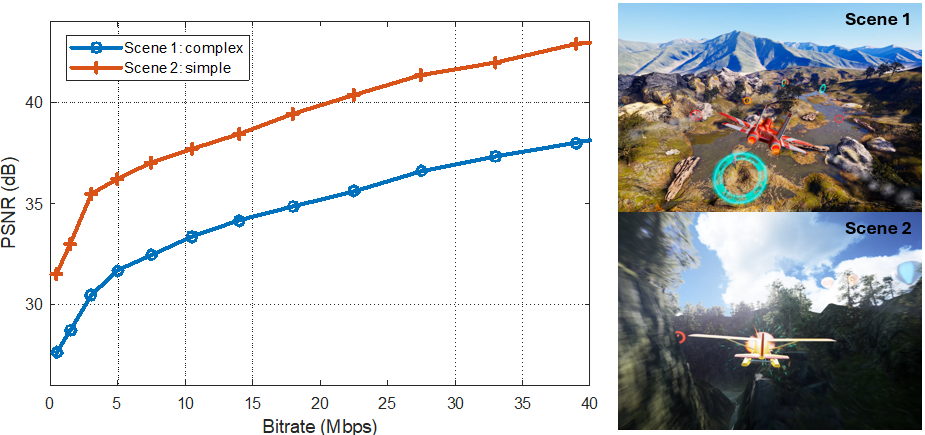}
\caption{Snapshots of different scenes of a cloud game and their RD (PSNR) curves. Scene 1 requires a bitrate of $\sim 19$ Mbps  to achieve a PSNR of 35 dB, while Scene 2 requires only $\sim 3$ Mbps to achieve the same PSNR value.}
\label{fig:scenes}
\end{center}
\vspace{-10pt}
\end{figure}

describing the quality degradation of the picture due to the combination of scene complexity and the artifacts of the compression/encoding process. Spatial quality metrics, such as Peak-Signal-to-Noise-Ratio (PSNR)  \cite{korhonen2012peak} and Video Multimethod Assessment Fusion (VMAF), compare the encoded video frame to the reference non-encoded frame on a pixel-level, block-level, or frame-level. Spatial quality can be measured and/or estimated by the AS during the frame encoding process, and is represented by a Rate-Distortion (RD) curve, which maps the encoding bitrate to the distortion (quality) of the frame.  An RD-curve depends on the complexity of the video scene, where more complex scenes (e.g., ones that are highly dynamic over time) require higher encoding bitrates to achieve the same quality as simple scenes (e.g., ones that are mostly static or slowly moving) encoded with a lower bitrate. For example, Fig.~\ref{fig:scenes} shows the RD curves of two scenes of a cloud game with varying degrees of complexity. Scene 1 (top right) is a complex scene that requires of  bitrate of $\sim 19$ Mbps  to achieve a PSNR of 35 dB, while Scene 2 (bottom right) requires $\sim 3$ Mbps to achieve the same PSNR value. RD curves of complex videos are typically steeper at higher bitrates than those of simpler videos. Moreover, the RD curve of a typical video stream does not remain constant all the time, and changes from one scene to another \cite{nadas2024qoe}. For an interactive XR application, video complexity changes drastically between instances of fast and slow head movement/rotation.

To unify both quality aspects, a Quality-Bitrate tradeoff curve (QB curve) is established for a specific scene using inputs from both the AS and AC, which maps the encoding bitrate to the overall achieved quality. For simplicity, in this paper, we target the temporal quality requirement through the rate allocation algorithm design, which we thoroughly discuss in Section \ref{sec:architecture}, and then we utilize only the PSNR RD curve from the AS as the QB curve of the transmitted video. More complex generation of QB curves from both the AS and AC inputs is part of future work. In this paper, we use PSNR and MSD as the quality metrics to define the UX of an XR device. More specifically, we define a satisfied UE as one whose PSNR is above a threshold $\gamma$ more than 95\% of the time, and whose MSD is less than $d_\text{stall}$.

\section{Proposed Framework for UX-aware Resource Allocation}
\label{sec:architecture}
\begin{figure}
\begin{center}
\includegraphics[width=0.98\linewidth]{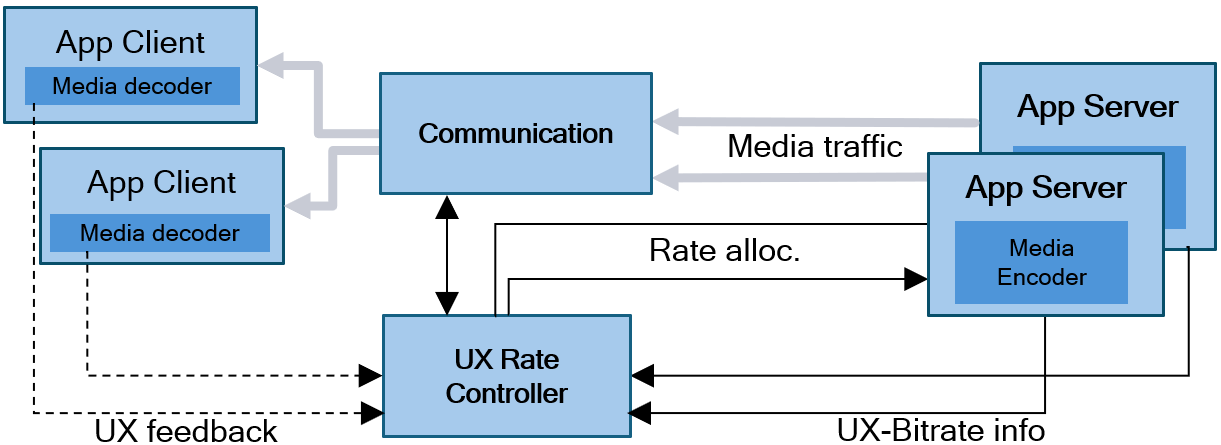}
\vspace{-1pt}
\caption{Proposed framework to UX-aware rate allocation.}
\label{fig:architecture}
\end{center}
\vspace{-10pt}
\end{figure}

As described previously, the video content complexity is an essential factor in determining the UX, and current cellular networks have no awareness of such complexity, which may deteriorate the overall experience of the UEs in the system.
In order to introduce UX-awareness, we propose to add a logical entity called \textit{UX rate controller} to the network, as shown in Fig. \ref{fig:architecture}. This controller receives updated video complexity information (in the form of updated QB curves) from the ASs. These updates can be configured to be periodic, or event-driven (i.e., update the QB curve upon significant video complexity change) \footnote{For most video streaming applications, QB curve update frequency can be in the order of hundreds of ms, or even seconds, making the proposed framework scalable. }. It also periodically receives updates from the network (e.g. Radio Access Network or RAN in 5G system) about the network conditions, e.g., the UEs' current SINR, MCS, or spectral efficiencies. Optionally, the controller may receive measurement feedback from the ACs about the current UX. The controller can then run rate allocation algorithms to maximize some UX-based network utility function, and communicate the allocated bitrates back to the ASs to encode their next frame(s). In the next subsections, we explore two possible examples for these optimizations: QoE-capacity maximization, and Max-min QoE fairness.

\subsection{QoE-capacity maximization (MaxCap)}
For this objective, the controller tries to maximize the network's QoE capacity, which is defined as the number of  satisfied UEs that the cell can simultaneously serve, with UE satisfaction as defined in Section \ref{sec:qoe_model}. The goal of the algorithm (which we summarize in Algorithm~\ref{alg:cap}) is to distribute the RAN resources in a specific duration $T_\text{win}^\text{qoe}$ among the UEs, with one unit of resource allocation being a Resource Block Group (RBG) (see Section 5.1.2.2 in~\cite{3gpp.38.214}). This resource allocation can then be translated into source bitrates depending on the channel qualities of the UEs. The outputs are then communicated with the ASs which adjust their encoding bitrates accordingly. Note that this increases the likelihood that the generated frames of all UEs can fit within the network's capacity, which will maintain the experienced delays of the frames low, making it more likely that the temporal quality requirements of the UX are met. The algorithm is periodically re-evaluated every $T_\text{period}^\text{qoe}$ to adapt to the current channel conditions and video complexities.

As inputs, the algorithm takes as input the total number of UEs in the cell $N_{\text{UE}}$, and the corresponding spectral efficiency of each UE $SE_n$ (where $n$ is the UE index) to capture the current channel conditions of the UEs, and the total number of available RBGs $N_\text{RBG}$ within the allocation period. Additionally, from each AS, the algorithm receives a target QoE value $\gamma_n$, as well as a QB curve  $Q_n(.)$, which is a function that maps the source bitrate to the achievable QoE. 

First, the algorithm calculates the achievable rate of UE~$n$ for each allocated RBG for that UE (line 4 in Algorithm  \ref{alg:cap}). This is dependent on the amount of app information bits (after accounting for header and signaling overheads) that the UE can fit within one RBG, which can be calculated as a function of $SE_n$ as described in 3GPP 38.214 \cite{3gpp.38.214}. The algorithm also calculates the minimum amount of RBGs needed for UE~$n$, $g_n$ (line 5 in Algorithm  \ref{alg:cap}), to meet its QoE target. If the total number of available RBGs is enough to accommodate the minimum amount of RBGs needed for all the UEs, then each UE is allocated its minimum amount of RBGs, and the rest of available RBGs are distributed equally among the UEs (lines 6-7). If that is not the case, some UEs will not be satisfied (i.e. will not meet their target QoE). To maximize the number of satisfied UEs, the UEs with the least amounts of needed RBGs to meet their QoE are admitted first. The remaining available resources are distributed equally among the unsatisfied UEs (lines 9-11). The final amount of allocated resource per UE is then used to calculate the achievable bitrate of that UE, and the achievable rates are communicated back to the servers.

\begin{figure}[!t]
\vspace{-8pt}
\noindent
\begin{minipage}{1\linewidth}
\begin{algorithm}[H]
\caption{Rate allocation for maximizing QoE capacity}\label{alg:cap}
\begin{algorithmic}[1]
\State\textbf{INPUTS (from network):} Number of UEs $N_{\text{UE}}$, Spectral efficiency per UE $SE_n\,\, \forall n$, Duration of resource allocation $T_\text{win}^\text{qoe}$.
\State\textbf{INPUTS (from AS(s)):}  QB function per UE $Q_n(.)\,\, \forall n$, QoE target per UE $\gamma_n\,\, \forall n$

\State\textbf{OUTPUTS (to AS(s)):} Allocated bitrate for each UE $R_n$

\State \textbf{Calculate} the total number of available RBGs  $N_\text{RBG}$ as the number of DL slots in $T_\text{win}^\text{qoe}$ multiplied by the number of RBGs per slot.

\State \textbf{Calculate} for each UE, the achievable rate per RBG as $R'_n =T(SE_n)/T_\text{win}^\text{qoe}$, where $T(SE_n)$ is the amount of app information bits per RBG and can be calculated using the formulas from chapter 5.1.3.2 in 3GPP~38.214~\cite{3gpp.38.214}.

\State \textbf{Calculate} for each UE, the minimum amount of RBGs needed to be satisfied $g_n = \lceil Q_n^{-1}(\gamma_n)/R'_n \rceil$
\If {$\sum_n g_n \le N_\text{RBG}$}
\State \textbf{Set} $R_n = \left( g_n + \lfloor \frac{N_\text{RBG} - \sum_n g_n}{N_\text{UE}} \rfloor \right) R'_n$
\Else
\State Sort UEs in ascending order of $g_n$, with new index $m$
\State Find the maximum number of satisfied UEs as $\max M \text{such that} \sum_{m=1}^M g_m < N_\text{RBG}$
\State \textbf{Set} 
\[ R_m = 
\begin{cases}
 g_mR'_m & \quad  \text{for} \,\,m \le M, \\
 \lfloor \frac{N_\text{RBG} - \sum_{m=1}^M g_m}{N_\text{UE} - M} \rfloor R'_m &\quad\text{for} \,\, m > M
\end{cases} 
\]

\EndIf
\end{algorithmic}
\end{algorithm}
\end{minipage}
\end{figure}

Note that the QoE-capacity maximization algorithm described above can be easily extended to handle other policies of dealing with unsatisfied UEs. For instance, the unsatisfied UEs can be downgraded to meet a lower QoE level (e.g. from excellent QoE to good or acceptable QoE) and the resources can be allocated to them accordingly. Also, Service Level Agreements (SLAs) can play a role in prioritizing the satisfaction of some UEs over others.

\subsection{QoE fairness (MaxMin)}
For this objective, the controller tries to maintain \textit{QoE fairness} among the UEs by maximizing the minimum QoE across the UEs in the cell. Similar to Algorithm \ref{alg:cap}, the maxmin fairness algorithm takes the same inputs and starts by calculating the achievable rate of UE~$n$ for each allocated RBG for that UE (see Algorithm \ref{alg:maxmin}). Then, the algorithm uses the well-known bisection method \cite{oliveira2020enhancement} to search for the rate allocation with which all the UEs in the cell can simultaneously maintain a maximum QoE value in the range $[Q_\text{min}, Q_\text{max}]$.

\begin{figure}[!t]
\vspace{-8pt}
\noindent
\begin{minipage}{1\linewidth}
\begin{algorithm}[H]
\caption{Rate allocation for maximizing minimum QoE}\label{alg:maxmin}
\begin{algorithmic}[1]
\State\textbf{INPUTS (from network):} Number of UEs $N_{\text{UE}}$, Spectral efficiency per UE $SE_n\,\, \forall n$, Duration of resource allocation $T_\text{win}^\text{qoe}$.
\State\textbf{INPUTS (from AS(s)):}  QB function per UE $Q_n(.)\,\, \forall n$

\State\textbf{OUTPUTS (to AS(s)):} Allocated bitrate for each UE $R_n$

\State \textbf{Calculate}  $N_\text{RBG}$ as the number of DL slots in $T_\text{win}^\text{qoe}$ multiplied by the number of RBGs per slot.
\State \textbf{Calculate} for each UE, the achievable rate per RBG as $R'_n =T(SE_n)/T_\text{win}^\text{qoe}$.

\State \textbf{Set} arbitrary $Q_\text{max}$ and $Q_\text{min}$

\While {$Q_\text{max} - Q_\text{min} > 0.5 $ dB}
\State Set  $Q_\text{mid}=\frac{Q_\text{max} + Q_\text{min}}{2}$
\State Find the minimum amount of resources needed for UE $n$ to maintain  $Q_\text{mid}$ quality,  $g_n = \lceil Q_n^{-1}(Q_\text{mid})/R'_n \rceil$

\If { $\sum_n g_n > N_\text{RBG}$ }
\State \textbf{Set} $Q_\text{max} = Q_\text{mid}$.
\ElsIf { $\sum_n g_n < N_\text{RBG}$ }
\State \textbf{Set} $Q_\text{min} = Q_\text{mid}$.
\Else
\State Break
\EndIf
\EndWhile
\State \textbf{Set} the bitrate for UE $n$ as  $R_n = g_n R_n'$.
\end{algorithmic}
\end{algorithm}
\end{minipage}
\end{figure}

It is worth noting that, while the concepts of this paper are developed for UEs with real-time media, they are generalizable to cases with mixed traffic. In such cases, each application may model the UX of its underlying traffic and shares its projected QoE as a function of bitrate with the UX rate controller. The UX rate controller may then assign bitrates to the different UEs (using the proposed algorithms) to satisfy their respective QoE requirements. Alternatively, each traffic type may be assigned a different priority level by RAN, and the proposed algorithms may then be used to allocate rates for UEs within each traffic priority. Other options for how to deal with mixed traffic scenarios is part of future investigation.

\section{Simulation Results}
\label{sec:results}
In this section, we present the performance evaluation results for our proposed UX-aware rate allocation algorithms. We first list our simulation parameters, describe the baseline algorithms against which we compare our proposed algorithms, and finally show the simulation results. 

\subsection{Simulation Parameters}
Table \ref{tab:sim_params} lists the simulation parameters for our performance evaluation platform. We first generate SINR traces for the UEs according to the 3GPP Indoor Hotspot (InH) and Urban Macro (UMa) channel models \cite{3gpp.38.901}. The InH channel model is applicable to VR scenarios, while the UMa channel model is applicable to AR scenarios. These result in a total of 33 cells (12 InH cells and 21 UMa cells) where the number of UEs per cell is swept from 1 to 10 UEs. The SINR traces are then used to simulate Over-The-Air (OTA) real-time 60-fps video transmission to the UEs. Each UE is sent a gaming video comprising of different scenes that vary in complexity, two of which are shown in Fig. \ref{fig:scenes}. The moments of switching between the scenes are randomized across the UEs.

\begin{table}
\caption{Simulation Parameters}
\label{tab:sim_params}
\vspace{-10pt}
\begin{center}
\begin{tabular}{|| c |  c | c ||} 
 \hline\hline
 \multirow{2}{*}{\textbf{Parameter}} & \multicolumn{2}{|c||}{\textbf{Value}} \\ 
\cline{2-3}
  & InH Channel & UMa Channel \\
 \hline\hline
  \multicolumn{3}{||c||}{\textbf{Network parameters}} \\
\hline
Carrier Frequency & 3.5 GHz & 4.7 GHz \\ 
\hline
 ISD (m) & 20 & 200  \\
\hline
 \# of gNBs & 12 & 7  \\
\hline
 \# of cells per gNB & 1 & 3  \\
\hline
Max gNB power & 23 dBm & 44 dBm  \\
\hline
Bandwidth (MHz) & \multicolumn{2}{|c||}{100 MHz (4 RBGs)} \\  
 \hline
 SCS  & \multicolumn{2}{|c||}{30 KHz} \\  
\hline
 Noise Figure  & \multicolumn{2}{|c||}{gNB: 5 dB, UE: 9 dB} \\  
\hline
 Scheduler & \multicolumn{2}{|c||}{Proportional Fair} \\  
\hline
Backhaul delay & \multicolumn{2}{|c||}{1 ms\footnotemark[1]} \\ 
\hline
 Target BLER  & \multicolumn{2}{|c||}{10\%} \\  
\hline
 Number of RBGs per slot  & \multicolumn{2}{|c||}{4} \\  
\hline
 Slot pattern  & \multicolumn{2}{|c||}{DDDSU} \\  
 \hline\hline
  \multicolumn{3}{||c||}{\textbf{Source parameters}} \\
\hline
Allowable source bitrates &  \multicolumn{2}{|c||}{1-50 Mbps} \\ 
\hline
Source fps &  \multicolumn{2}{|c||}{60} \\ 
 \hline
Average scene duration &  \multicolumn{2}{|c||}{3.5 seconds} \\ 
 \hline
Encoding delay  &  \multicolumn{2}{|c||}{1 ms} \\ 
\hline
Decoding delay  &  \multicolumn{2}{|c||}{1 ms} \\ 
\hline\hline
  \multicolumn{3}{||c||}{\textbf{UX-aware rate allocation algorithms parameters}} \\
\hline
$T_\text{win}^{qoe}$&  \multicolumn{2}{|c||}{15 ms} \\ 
\hline
 $T_\text{period}^\text{qoe}$ &  \multicolumn{2}{|c||}{33 ms} \\ 
\hline
 QoE target $\gamma$ &  \multicolumn{2}{|c||}{35 dB PSNR} \\ 
\hline
Max stall duration ($d_\text{stall}$) &  \multicolumn{2}{|c||}{50 ms} \\ 
\hline
$Q_\text{min}, Q_\text{max}$ (maxmin alg.) &  \multicolumn{2}{|c||}{30 dB, 40 dB PSNR} \\
\hline\hline
   \multicolumn{3}{||c||}{ \textbf{RTT-based rate control algorithm parameters}} \\
\hline
 $T_\text{period}^\text{RTT}$ &  \multicolumn{2}{|c||}{50 ms} \\ 
 \hline
$T_\text{win}^\text{RTT}$ &  \multicolumn{2}{|c||}{100 ms} \\ 
\hline
$\alpha_\text{up},\, \alpha_\text{down}$ &  \multicolumn{2}{|c||}{1.1, 0.9} \\ 
 \hline
$\beta_\text{low}^\text{RTT},\,\beta_\text{high}^\text{RTT}$ &  \multicolumn{2}{|c||}{8 ms, 10 ms} \\ 
\hline\hline
 \multicolumn{3}{||c||}{\textbf{L4S framework parameters}} \\
\hline
 $\beta_\text{low}^\text{L4S}$ &  \multicolumn{2}{|c||}{4 ms} \\ 
\hline
$\beta_\text{high}^\text{L4S}$ &  \multicolumn{2}{|c||}{17 ms} \\ 
\hline\hline
\end{tabular}
\end{center}
\vspace{-10pt}
\end{table}

We compare the performance of our proposed UX-aware rate allocation algorithms to two conventional rate control algorithms, which can be broadly classified into Over-The-Top (OTT) algorithms, and network-assisted algorithms.

\subsubsection{RTT-based Rate Control} 
In this simple OTT algorithm, the AS initializes its bitrate randomly between 1 and 50 Mbps. The application client at the UE sends a feedback report every $T_\text{period}^\text{RTT}$ ms which includes the average measured RTT within a window of duration $T_\text{win}^\text{RTT}$ ms. Upon the reception of the report, the AS increases its current bitrate by a multiplicative factor of $\alpha_\text{up}$ if the average RTT is smaller than  $\beta_\text{low}^{RTT}$ ms, and decreases its current bitrate by a multiplicative factor of $\alpha_\text{down}$ if the average RTT is greater than $\beta_\text{high}^{RTT}$ ms, and keeps the current bitrate unchanged otherwise. Similar RTT-based rate control algorithms have been proposed in the literature \cite{maura2024experimenting}.

\subsubsection{Prague Congestion Control} Low Latency, Low Loss, and Scalable Throughput (L4S) is one example of network-assisted frameworks which is standardized by IETF in RFC~9330 \cite{rfc9330}. A network node (e.g., RAN) marks the IP packets using the Explicit Congestion Notification (ECN) field in the IP packet header, with a marking probability that is an increasing function of the queueing delay experienced at the node. The marking policy in our implementation is to have a zero marking probability for queueing delay $\le \beta_\text{low}^{L4S}$ ms, a 100\% marking probability for delays $\ge \beta_\text{high}^{L4S}$ ms, and linear in between. Finally, an L4S-compliant end-to-end rate adaptation algorithm utilizes these markings to adjust the source bitrate. Prague Congestion Control \cite{briscoe2019implementing} is one such rate adaptation algorithm that we utilize as a baseline for this study, and is characterized by: 1) additive bitrate increase for every unmarked packet, 2) multiplicative decrease (once per RTT) for marked packets by a factor of $(1-m_\text{ecn}/2)$, where $m_\text{ecn}$ is the fraction of recently marked packets, and 3) multiplicative decrease upon packet loss by a factor of 1/2.

The parameter values of the baseline algorithms used in our study are provided in Table \ref{tab:sim_params}.
\footnotetext[1]{ Small backhaul delay due to the assumption of colocation of the application server with 5G system.}

\begin{figure}
\begin{centering}
\includegraphics[width=1\linewidth]{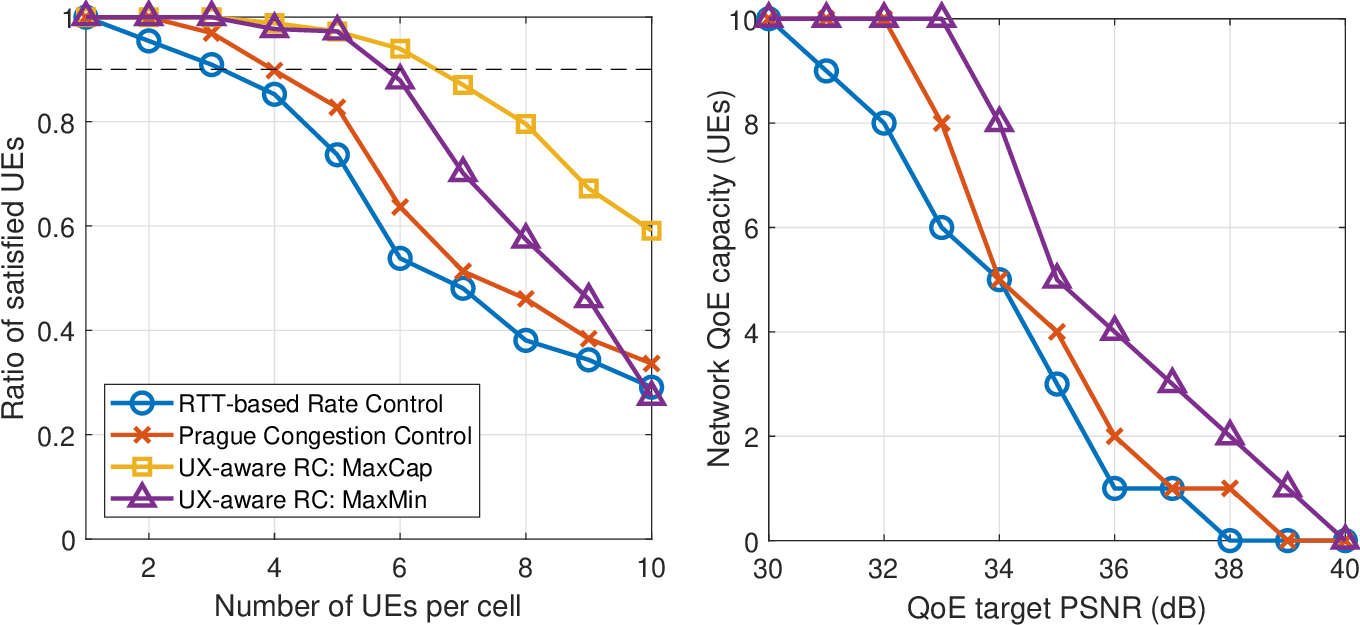}
\vspace{-10pt}
\caption{(a) Ratio of satisfied UEs as a function of the number of UEs per cell. (b) Network QoE capacity as a function of the target QoE threshold $\gamma$.}
\label{fig:QoEcapacity}
\end{centering}
\vspace{-10pt}
\end{figure}

\subsection{Simulation Results}
Fig. \ref{fig:QoEcapacity} (a) shows the UE satisfaction rate as a function of the number of UEs per cell. A satisfied UE is one whose PSNR is above a threshold $\gamma$ more than 95\% of the time, and whose maximum stall duration is less than $d_\text{stall}$. At 6 UEs per cell, it can be seen that the ratio of satisfied UEs is 93.9\% with the MaxCap algorithm, and 87.8\% with the MaxMin algorithm, compared to 63.6\% satisfaction ratio with Prague congestion control, and 53.8\% satisfaction ratio with RTT-based rate control. Following similar definitions of XR capacity in 3GPP \cite{3gpp.38.835}, we define QoE capacity as the maximum number of UEs per cell, where at least 90\% of the UEs in that cell are satisfied.  It can be seen from Fig. \ref{fig:QoEcapacity} that the QoE capacity of the proposed MaxCap algorithm is~6, and the MaxMin algorithm is~5, while that of Prague congestion control is~4 and RTT-based rate control is~3, showing that UX-aware rate control provides a 50\%---100\% QoE capacity gain when compared to conventional rate control algorithms.

While the MaxMin algorithm is not designed to maximize the network QoE capacity given a specific target QoE threshold, Fig.~\ref{fig:QoEcapacity} (b) shows that it consistently outperforms the conventional rate control algorithms over the range of possible QoE PSNR target thresholds $\gamma$, since it aims at converging to an operating point where all UEs have the same maximum possible QoE.

\begin{figure}
\includegraphics[width=1\linewidth]{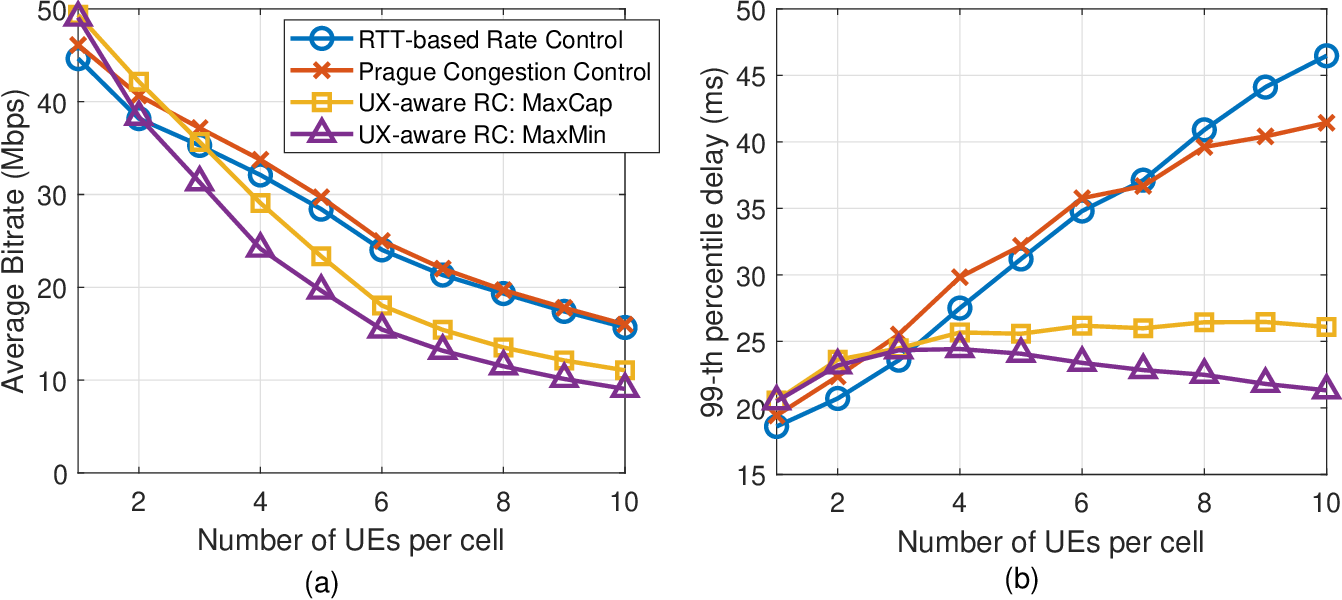}
\vspace{-15pt}
\caption{(a) Average source bitrate and (b) $99^\text{th}$ percentile frame delay, as a function of the number of UEs per cell. }
\label{fig:AvgBitrates}
\vspace{-12pt}
\end{figure}

When examining the average source bitrates of the UEs, as shown in Fig. \ref{fig:AvgBitrates}~(a), it can be seen that UX-aware rate control achieves higher QoE gains while maintaining the average bitrate lower than the conventional rate control algorithms. This is due to the fact the UX-aware allocation limits/caps the bitrate of the UEs with simple scenes and/or very good channel conditions, which would otherwise have unnecessarily increased their bitrate considerably. Moreover, Fig. \ref{fig:AvgBitrates}~(b) shows that UX-aware rate control achieves a much lower 99$^\text{th}$ percentile frame delay compared to conventional rate control algorithms, which positively impacts the temporal quality aspect of the UX, since stall durations are function of the frame latency.
The latency reduction is due to: 1) the UX-aware rate allocation algorithm design which tries to fit the bitrates of the UEs within the network's capacity, as explained in Section~\ref{sec:architecture}, 2) the overall decrease in average bitrate  achieved by the UX-awareness at the network, and 3) the fast response of our proposed framework to sudden channel variations using the direct feedback to the server. 

\begin{figure}
\includegraphics[width=0.97\linewidth]{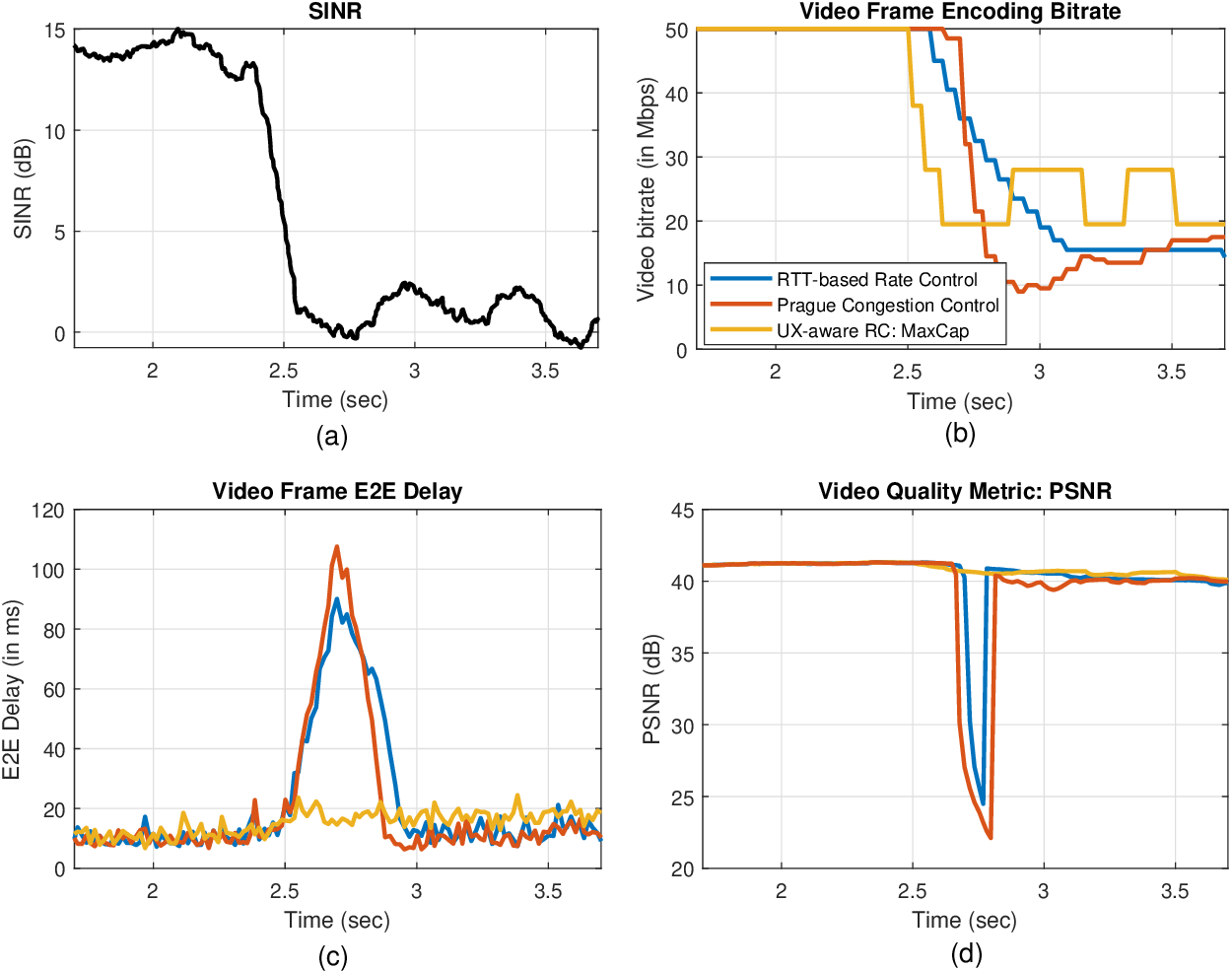}
\vspace{-8pt}
\caption{Comparison of the performance of the proposed UX-aware rate allocation and the baseline algorithms in response to sudden channel variations. See the colored PDF version for optimal viewing of this figure.}
\label{fig:single_ue}
\vspace{-14pt}
\end{figure}

To verify the last point, we run a single UE simulation where the SINR trace drops abruptly, e.g., due to sudden blockage and/or interference, see Fig. \ref{fig:single_ue} (a). Fig. \ref{fig:single_ue} (b) shows the adapted source bitrate of the different rate control algorithms in response to the channel variation. It can be seen that the baseline algorithms take more time to adapt to the new channel condition. During this transition period, the end-to-end delay of several frames becomes very high due to the queue accumulation at the gNB in the baseline algorithms, see Fig. \ref{fig:single_ue} (c), which results in some of these frames being lost/dropped at the UE, and the UE entering a stall that negatively impacts its UX, as can be seen in Fig.~\ref{fig:single_ue}~(d). Our proposed UX-aware rate control algorithm does not suffer from such drawbacks.

\section{Conclusions and Future Work}
\label{sec:conclusion}
In this paper, we proposed a framework for communication networks to provide UX guarantees to its users by requiring the application servers to share real-time media complexity information with the network. We demonstrated the potential benefits of this UX-awareness at the network by introducing two different rate allocation algorithms that maximize the network's QoE capacity, and the network's QoE fairness (in a maxmin sense), respectively. Our simulation results show that this framework can achieve $\sim$50\%---100\% gain in the network's QoE capacity when compared to conventional rate control algorithms. At the same time, our proposed framework is shown to reduce the overall average bitrate of the UEs as well as the latency of the video frame delivery. 

Some issues remain open and require further studying as part of future work. For instance, methods for real-time estimation of RD-curves at the video encoders need to be designed, and the impact of imperfect RD-curve estimation on the overall performance of the algorithms needs to be assessed.

\bibliographystyle{ieeetr}
\bibliography{refs.bib}

\end{document}